\documentclass[%
 aip,
 amsmath,amssymb,
 reprint,%
]{revtex4-2}

\usepackage{graphicx}% Include figure files
\usepackage{dcolumn}% Align table columns on decimal point
\usepackage{bm}% bold math
\usepackage{amsmath}
\usepackage[utf8]{inputenc}
\usepackage[T1]{fontenc}
\usepackage{mathptmx}
\usepackage{xcolor}
\usepackage{braket}

\begin{document}

%\preprint{}

\title{Phase-Stable Optical Fiber Links for Quantum Network Protocols}

%Precision metrology made easy: a guide to the Er/Yb:glass optical frequency comb



\author{N. V. Nardelli}
 \email[Author to whom correspondence should be addressed: ]{nicknardelli@gmail.com.}
  \affiliation{ 
National Institute of Standards \& Technology, 325 Broadway, Boulder, CO 80305, USA%\\This line break forced with \textbackslash\textbackslash
}%

\author{D. V. Reddy}
\affiliation{ 
National Institute of Standards \& Technology, 325 Broadway, Boulder, CO 80305, USA%\\This line break forced with \textbackslash\textbackslash
}
\affiliation{%
University of Colorado, Boulder, Colorado 80309, USA%\\This line break forced% with \\
}%

\author{M. Grayson}
\affiliation{ 
National Institute of Standards \& Technology, 325 Broadway, Boulder, CO 80305, USA%\\This line break forced with \textbackslash\textbackslash
}

\author{D. Sorensen}
\affiliation{ 
National Institute of Standards \& Technology, 325 Broadway, Boulder, CO 80305, USA%\\This line break forced with \textbackslash\textbackslash
}
\affiliation{%
University of Colorado, Boulder, Colorado 80309, USA%\\This line break forced% with \\
}%

\author{M. J. Stevens}
\affiliation{ 
National Institute of Standards \& Technology, 325 Broadway, Boulder, CO 80305, USA%\\This line break forced with \textbackslash\textbackslash
}

\author{M. D. Mazurek}
\affiliation{ 
National Institute of Standards \& Technology, 325 Broadway, Boulder, CO 80305, USA%\\This line break forced with \textbackslash\textbackslash
}
\affiliation{%
University of Colorado, Boulder, Colorado 80309, USA%\\This line break forced% with \\
}%

\author{L. K. Shalm}
\affiliation{ 
National Institute of Standards \& Technology, 325 Broadway, Boulder, CO 80305, USA%\\This line break forced with \textbackslash\textbackslash
}
\affiliation{%
Quantum Engineering Initiative, University of Colorado, Boulder, Colorado 80309, USA%\\This line break forced% with \\
}%

\author{T. M. Fortier}%
\affiliation{ 
National Institute of Standards \& Technology, 325 Broadway, Boulder, CO 80305, USA%\\This line break forced with \textbackslash\textbackslash
}%


\date{\today}

% approximately 250 words
\begin{abstract}
We demonstrate the distribution of single-photon-level pulses from a mode-locked laser source over a phase-stable fiber link, achieving an optical timing jitter of less than 100 as over 10 minutes of data accumulation. This stability enables a fidelity greater than 0.998 between two stabilized 2.1 km long deployed fiber links. Building on time and frequency metrology techniques traditionally used for high-stability optical atomic clock signal distribution, we use time and frequency multiplexing to achieve an isolation of quantum and classical channels of $8 \times 10^{10}$. Our results mark a necessary step towards scalable, high-rate quantum networks with a provable quantum advantage.
\end{abstract}


\maketitle %\maketitle must follow title, authors, abstract and \pacs

%\tableofcontents

%%%%%%%%%%%%%%%  Main text   %%%%%%%%%%%%%%%
% \linenumbers
%%%%%%%%%%%%%%%%%%%%%%%%%%%%%%%%%%%%%%%%%%%%%%%%%%%%%%%%%%%% 
%%%%%%%%%%%%%%%%%%%%%%%%%%%%%%%%%%%%%%%%%%%%%%%%%%%%%%%%%%%% 
%%%%%%%%%%%%%%%%%%%%%%%%%%%%%%%%%%%%%%%%%%%%%%%%%%%%%%%%%%%% 
\section{Introduction}
The future success of many quantum technologies, including computing \cite{PhysRevA.76.062323}, communications \cite{RevModPhys.74.145,Lucamarini2018,Zhou2023}, sensing, and metrology \cite{Gottesman2012,Czupryniak2023}, hinges on the ability to reliably distribute quantum entanglement across a network. Because quantum states cannot be copied, transmission loss poses a fundamental challenge for long-distance quantum communication. Overcoming this limitation requires quantum repeaters, which enable efficient distribution of entanglement over lossy channels. The first quantum repeater architecture, proposed by Duan, Lukin, Cirac, and Zoller (DLCZ) \cite{DLCZ2001}, relies on quantum interference between two indistinguishable pathways corresponding to the emission of a single photon from spatially separated atomic ensembles. After interfering these two paths, the detection of a single photon heralds the creation of an entangled state between the ensembles (as shown in Fig. \ref{Fig:DLCZ} a)), provided that the photonic paths are indistinguishable. A central challenge for DLCZ-based architectures is their sensitivity to optical phase fluctuations. Because the distributed entangled state depends critically on the relative phase between the photonic paths, all fiber links in the network effectively form a large interferometer. Path-length fluctuations on the order of tens of nanometers are sufficient to wash out interference and significantly reduce entanglement fidelity. Entanglement purification \cite{PhysRevA.59.169, PhysRevLett.81.5932, RevModPhys.83.33} can be used to create higher fidelity entanglement from multiple lower fidelity entangled states. However, the number of required copies increases as the initial entangled state fidelity drops. Consequently, even modest reductions in phase-noise-induced infidelity can substantially enhance repeater throughput and link distance.

In this work we demonstrate a telecom-wavelength, multi-fiber, quantum-compatible phase-stabilization technique for DLCZ-style quantum repeaters that performs nearly an order of magnitude better in interference visibility compared to previously reported approaches. Using methods adapted from the atomic clock and optical time-transfer communities, we independently stabilize two 2.1 km single-mode fibers in the interferometric configuration required for entanglement swapping in a quantum repeater. Excess noise was intentionally introduced by aerially suspending the fibers, in contrast to equivalent underground deployments, providing a stringent test of stabilization performance under realistic environmental perturbations. In this deployed-fiber demonstration, we achieve three critical performance benchmarks for path-entangled quantum networking: (1) optical RMS timing jitter around 70 as over 20 seconds (corresponding to a global phase noise of $\Delta \phi_\mathrm{RMS} = 0.085$ radians) and a timing drift of less than 100 as over 10 minutes, (2) path indistinguishability exceeding 99.6 $\%$, and (3) isolation between classical stabilization signals and the quantum channel greater than $8 \times 10^{10}$, defined as the ratio of photons in the classical channel to those coupling into the quantum channel. Our approach is suitable for both matter-based and high event rate all-optical DLCZ repeater architectures needed to support municipal-scale distributed quantum applications.

A key advantage of the DLCZ architecture lies in its scaling behavior: repeater protocols based on two-photon encodings (e.g., polarization or frequency) typically exhibit entanglement rates that scale linearly with the transmission efficiency, $\eta$, whereas DLCZ-type protocols retain a $\sqrt\eta$ scaling\cite{Stolk2025}. This more favorable scaling makes DLCZ-style repeaters particularly attractive as network length and complexity increase. As a result, state-of-the-art quantum repeater demonstrations \cite{Liu2024, Stolk2024} are based on variants of the DLCZ protocol. However, simultaneously stabilizing multiple deployed fibers is challenging and typically requires using bright classical light to measure relative path length changes. However, this bright light must coexist with the single-photons used to perform the entanglement swapping. To prevent unwanted crosstalk between the classical and quantum channels, a high degree of isolation is required. 

Fluctuations or drifts in the relative phase between different paths place a fundamental upper bound on the achievable entanglement fidelity, $F$. For a path-entangled state, this bound is given by
\begin{equation}\label{Eq:fidelity}
F(\Delta\phi_{\mathrm{RMS}}) = \left| \langle \psi_\text{path} | \psi_\text{path}' \rangle \right|^2
= \frac{1}{2}\left(1 + \cos(\Delta\phi_{\mathrm{RMS}})\right),
\end{equation}
where $\Delta \phi_\mathrm{RMS}$ is the root-mean-square phase error. Recent experimental demonstrations highlight the impact of this limitation. In ref. \cite{Liu2024}, the repeater links exhibited an RMS phase error of 0.297 radians, corresponding to a maximum achievable fidelity of $\sim 0.98$. Similarly, ref. \cite{Stolk2025} reported an RMS phase error of 0.609 radians, limiting the fidelity to $\sim 0.91$. The stabilization requirements for a phase-sensitive quantum repeater are also subtly different than those needed for other quantum communication protocols, such as twin-field quantum key distribution (TF-QKD)\cite{Lucamarini2018,Zhou2023}. In recent TF-QKD demonstrations the light in the quantum channel is bright enough to use to directly track the phase, removing the need for a separate stabilization light source. Additionally, the needed corrections can be done via post processing. In contrast, for general purpose DLCZ-style repeaters phase fluctuations must be dynamically corrected either by dynamically adjusting the path lengths involved or by applying corrections to the stored qubits at the end stations.

%%%%%%%%%%%%%%%%%%%%%%%%%%%%%%%%%%%%%%%%%%%%%%%%%%%%%%%%%%%% 
%%%%%%%%%%%%%%%%%%%%%%%%%%%%%%%%%%%%%%%%%%%%%%%%%%%%%%%%%%%% 
%%%%%%%%%%%%%%%%%%%%%%%%%%%%%%%%%%%%%%%%%%%%%%%%%%%%%%%%%%%% 
\section{All-optical DLCZ architecture}
Although most quantum repeater architectures rely on matter-based qubits at the end stations, an all-optical implementation of the DLCZ repeater is possible. In this approach, matter-based qubits are replaced by on-demand single-photon sources and linear optical elements, with entanglement encoded in the photonic path degrees of freedom. In principle, single-photon sources can operate at gigahertz rates, leading to substantially higher entanglement swapping event rates. Available all-optical quantum memories\cite{Kaneda:17} mean that this architecture is best suited for high-performance metropolitan-scale quantum networks. The techniques developed in this work show that optical paths in a high-bandwidth, all-optical DLCZ-based quantum repeater can be stabilized with high visibility and strong isolation between classical and quantum channels. This represents a critical first step toward building scalable, high-performance, all-photonic entanglement swapping in metropolitan-scale networks.

\label{sec:network}
\begin{figure}
  \centering
  \includegraphics[width=8.5cm]{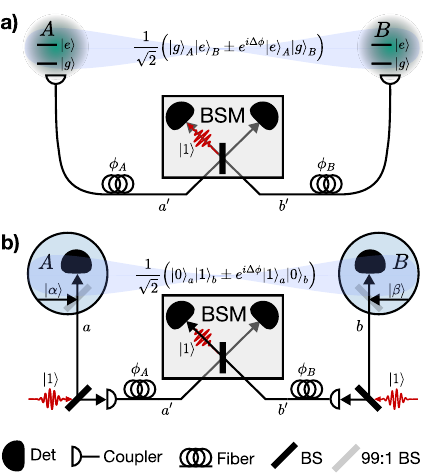}
  \caption{a) Simplified DLZC scheme entangling two remote atomic qubits. Initially, each atom is prepared in an excited state with a probability of decaying to the ground state and emitting a photon. Emitted photons from either atom are coupled into fiber and sent to a Bell State Measurement (BSM) comprised of a 50/50 beamsplitter (BS) that erases which-path information followed by single-photon detectors at either exit port. When a single photon is detected in the BSM, it is unknown which atom $A$ or $B$ is in the ground state, resulting in the entangled Bell State between atoms $A$ and $B$. Here the relative phase $\Delta\phi$ depends on the difference in phase due to path length differences between $\phi_{A}$ and $\phi_{B}$. b) In the all-optical version of DLCZ, the role of the atom is replaced by a single-photon source and beamsplitter. This initially creates entanglement between the paths $a$ and $a'$ at node $A$ and entanglement between the paths $b$ and $b'$ at node $B$.  Modes $a'$ and $b'$ are sent to a remote BSM station where the detection of a single-photon leaves modes $a$ and $b$ entangled with one another. The entanglement at nodes $A$ and $B$ can be measured using independent weak-local oscillators $\ket{\alpha}$ and $\ket{\beta}$ as described in the text.}
  \label{Fig:DLCZ}
\end{figure}

The proposed all-optical DLCZ-based quantum networks is depicted in Fig. \ref{Fig:DLCZ} b). In such networks, the role of each atomic ensemble is replaced by a heralded single-photon source and a beamsplitter. Consider a quantum repeater consisting of two remote locations $A$ and $B$. After passing through a 50/50 beamsplitter, the single photon at node $A$ ($B$) is transformed into the path-entangled state $\frac{1}{\sqrt{2}}\left(\ket{0}_{a (b)}\ket{1}_{a' (b')}+ \ket{1}_{a (b)}\ket{0}_{a' (b')}\right)$, where $a$ ($b$) and $a'$ ($b'$) are the output modes of the beamsplitter, $\ket{1}$ is a single photon state, and $\ket{0}$ is the vacuum state. This is analogous to the entangled state generated between a two-level system $A$ ($B$) and a photon emitted into mode $a'$ ($b'$) at the start of the DLCZ protocol, $\frac{1}{\sqrt{2}}\left(\ket{g}_{A(B)}\ket{1}_{a' (b')}+ \ket{e}_{A (B)}\ket{0}_{a' (b')}\right)$. One path at each node is stored locally ($a$ and $b$) while the other path is sent to an entanglement swapping station ($a'$ and $b'$) where a Bell state measurement (BSM) is performed using linear optics as shown in Fig. \ref{Fig:DLCZ}. The two paths from the remote nodes are interfered on a beamsplitter. If a single photon is detected at an output port of the beamsplitter, this heralds the successful entanglement of the two local paths ($A$ and $B$) stored at the remote nodes\cite{PhysRevA.76.050301, PhysRevA.90.033836, riedmantten2021}:
\begin{equation} \label{Eq:quantum state}
    | \psi_\text{path} \rangle = \frac{1}{\sqrt{2}} \left( |0\rangle_a |1\rangle_b \pm e^{i\Delta\phi} |1\rangle_a |0\rangle_b \right).
\end{equation}
The phase term, $\Delta \phi$, represents the difference in accumulated optical phase between the two paths $a'$ and $b'$ with a length difference of $\Delta L$. The relationship between phase and length, $\Delta\phi = k \times \Delta L$, where $k$ indicates the wavevector, demonstrates the sensitivity of the quantum state to perturbations in optical path length. Therefore, to distribute path entanglement over optical fiber links with high fidelity, it is necessary to actively stabilize the fiber length. An additional overall $\pi$ phase shift is introduced depending on which detector in the BSM fires. Using feed forward, this extra phase shift can be corrected. 

The two remote nodes can in principle couple each of their entangled paths to an atomic ensemble to create entanglement between their respective ground and excited states. Alternatively, the remote nodes can either store their quantum states in a phase-stable variable optical delay that acts as a memory, or they can directly measure their path entangled qubit using a weak local oscillator as shown in Figure \ref{Fig:DLCZ} b). A weak local oscillator and single-photon detector can approximate single-qubit measurements in a path-entangled encoding\cite{PARIS199678, PhysRevLett.82.2009}. A weak local oscillator, for instance, can perform the projective measurements needed to show a violation of a Bell inequality needed for advanced device-independent quantum communication protocols as shown in \cite{PhysRevLett.82.2009, PhysRevA.88.012111, PhysRevA.76.052101}. The benefit of an all-optical DLCZ repeater is that the heralded photon source may operate at trial rates orders of magnitude higher than comparable matter-based systems, but comes with the additional requirement that the local oscillators at each location share the same phase.

\begin{figure*}[!t]
  \centering
  \includegraphics[width=17cm]{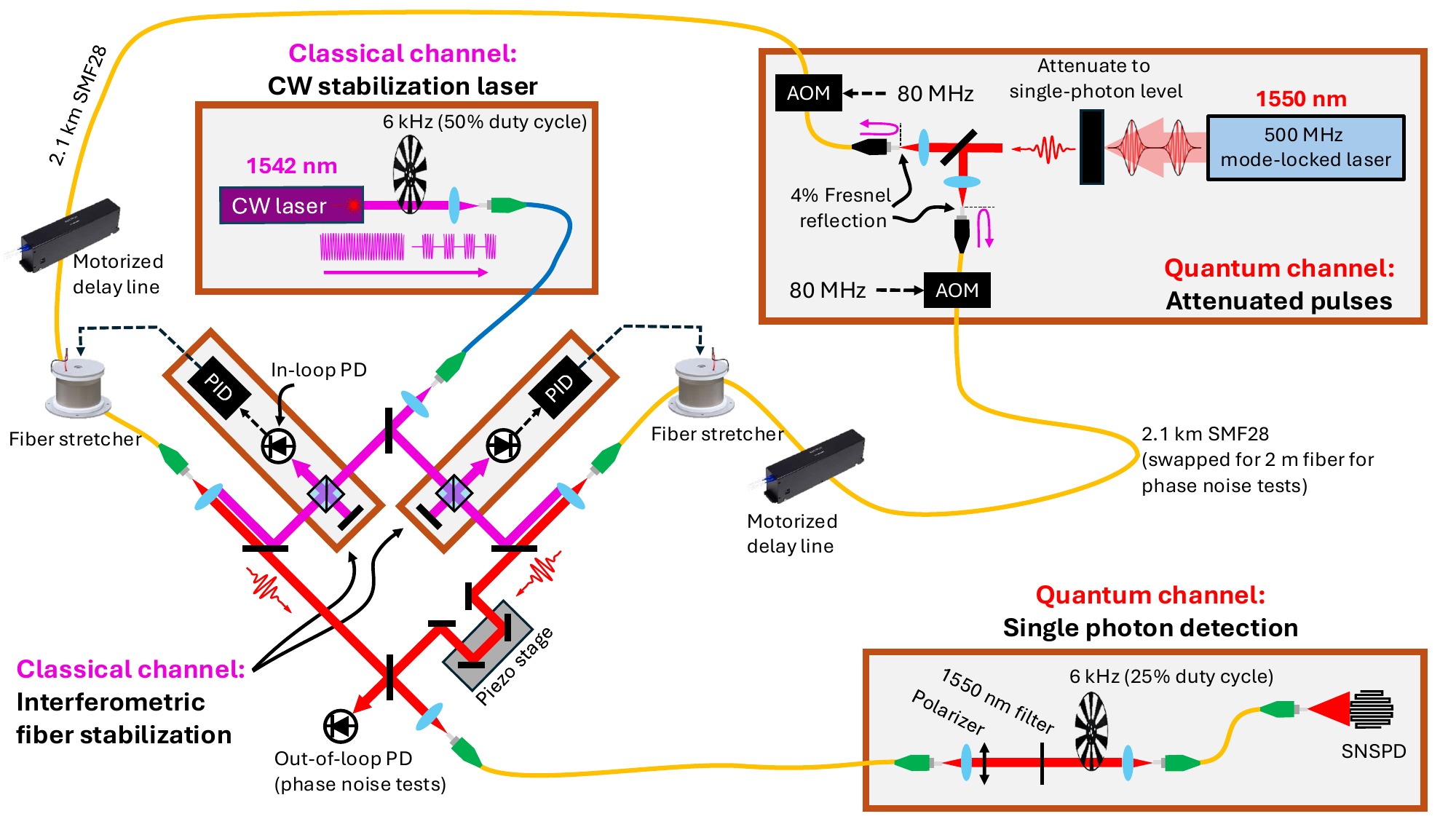}
  \caption{Experimental setup used for the phase stabilization and noise measurement of a deployed 2.1 km fiber link. This setup operates by counter-propagating light from the single-photon quantum channel, in the same spatial mode as classical single-frequency light from the phase-stabilization channel. The channels are isolated from one another by time-multiplexing the classical and quantum channels using a mechanical chopper. The quantum signal, generated by attenuating a 500 MHz repetition rate, 1550 nm mode-locked laser to an average of $<1$ photon per pulse, is combined with the classical channel via a 50/50 optical beamsplitter. Length compensation is achieved using round trip measurement of the fiber phase noise, used in a feedback loop, and actuated via a fiber stretcher and motorized delay line.  Two independently stabilized 2.1 km fiber links are employed to assess the stabilized performance of both channels by measuring the phase noise power spectral density of the quantum channel. A third short link is used to connect the output of the free space interferometer with a superconducting nanowire single-photon detector (SNSPD) to assess the fringe visibility of single photons traversing the stabilized fiber link. In the figure, AOM - acousto-optic modulator, PID - proportional/integral/derivative loop filter.}
  \label{fig:setup}
\end{figure*}

To achieve high-fidelity (> 0.99) path-state entanglement distribution, several conditions must be met. Phase stability is crucial, as even small fluctuations in optical path length can introduce phase noise that limits the measurement fidelity. A phase change of $\Delta\phi = 0.1$ radians reduces the fidelity to about 0.9975 (discussed in more detail in section \ref{sec3b}). Additionally, the optical paths from source to measurement must be highly indistinguishable, that is, there must be no identifying information about which source a photon came from that reduces the interference visibility at the repeater. Photons from the two sources must have the same polarization, spectral and temporal profiles,  and arrive at the same time within femtoseconds. Finally, a high degree of isolation between quantum and classical channels is required to ensure that the relatively high-power stabilization laser does not overwhelm the single-photon detectors or lower the fidelity with noise photons. Implementing a full entanglement swapping demonstration is beyond the scope of this manuscript. Instead, we focus on independently stabilizing the two deployed fiber paths that would connect nodes A and B to the swapping station in a Mach-Zehnder interferometer (as shown in Fig. \ref{fig:setup}) that would enable such a demonstration. We use an attenuated laser at 1550 nm to simulate a heralded telecom photon source operating with a bandwidth of 1 nm at a repetition rate of 500 MHz and average photon number of 0.08. A major benefit of our scheme is that the stabilization signal that is distributed can in principle be used to lock independent weak local oscillators needed to carry out the measurements on the path-entangled qubits.

%%%%%%%%%%%%%%%%%%%%%%%%%%%%%%%%%%%%%%%%%%%%%%%%%%%%%%%%%%%% 
%%%%%%%%%%%%%%%%%%%%%%%%%%%%%%%%%%%%%%%%%%%%%%%%%%%%%%%%%%%% 
%%%%%%%%%%%%%%%%%%%%%%%%%%%%%%%%%%%%%%%%%%%%%%%%%%%%%%%%%%%% 
\section{The classical channel}
\label{sec:classical}

%%%%%%%%%%%%%%%%%%%%%%%%%%%%%%%%%%%%%%%%%%%%%%%%%%%%%
\subsection{Optical Phase Stabilization}
To achieve the desired level of phase stability, we employ an experimental technique that closely parallels a fiber stabilization architecture used in the distribution of high-stability optical signals from optical atomic clocks \cite{Ma1994}, see Fig. \ref{fig:setup}. These fiber links enable relative frequency comparisons of remote optical clocks at the $10^{-18}$ level \cite{BACON2021}.

A key difference between stabilization schemes is that in atomic clock networks the light used to stabilize the network link is also the light that must be transmitted. In the current demonstration, the classical signal used for stabilization is separate from the quantum signal. However, to achieve the best stabilization performance, the classical channel must sample the same fiber length and refractive index fluctuations as the quantum channel. As a result, it must co-propagate with the quantum signal, sharing the same optical frequency and polarization. One of the challenges in this architecture is isolating the classical and quantum signals from one another. To achieve this, we use mechanical choppers to time-multiplex the classical stabilization light and the single-photon-level quantum signals. Additionally, we sacrifice some channel commonality by operating them about 8 nm apart, with the classical channel at 1542 nm and the quantum channel at 1550 nm. This allows for further isolation via wavelength division multiplexing.

The classical channel consists of a narrow-linewidth continuous wave (CW) laser at 1542 nm, stabilized to an optical reference cavity that enables a full-width half-maximum linewidth, $\Delta \nu$, less than 10 mHz  \cite{Matei2017}. The choice of stabilization laser is important since the coherence time of the laser, $\tau_\text{coh} = 1/(\pi \Delta \nu)$, must far exceed the transit time of light through the optical fiber. If the laser phase drifts over the fiber transit time, the system cannot separate laser noise from fiber noise, causing the laser’s phase fluctuations to be imprinted onto the link instability.

The CW light is chopped by a 6 kHz chopper with 50\% duty cycle, which is anti-synchronized with a 25\% duty cycle chopper located before the single photon detector in the quantum channel. We use mechanical choppers because they achieve theoretically perfect extinction while other modulators (e.g., EOMs or AOMs) can achieve only up to 60 dB extinction under very carefully controlled conditions. After chopping, a fraction of the CW light is split off by a beam splitter and used as a stable interferometric reference, while the remainder of the light traverses the optical fiber. At the remote end, about 4\% (Fresnel reflection) of the signal is retro-reflected from the flat fiber tip such that it returns through the fiber and interferes with the reference light at the local end, generating an error signal that is used to feedback and suppress the fiber noise. An acousto-optic modulator (AOM) is located at the remote end of the fiber, which shifts the reflected optical frequency by twice the RF drive frequency (80 MHz). The AOM shifts the interference signal frequency away from DC, which avoids DC noise sources such as photodetector flicker noise and renders the stabilization immune to laser power fluctuations. Additionally, placing the AOM at the end of the link frequency-shifts the returning light, allowing reflections from the fiber end to be distinguished from unwanted reflections within the link. This separation improves the signal-to-noise ratio of the stabilization signal \cite{Williams2008}.

A fiber stretcher serves as the fast feedback actuator (>20 kHz bandwidth), while a motorized delay line provides slow, large-range path-length adjustments. It is essential that the actuator modify the optical path length rather than only the optical phase (as would an acousto-optic modulator), because the stabilization must control both the phase and the arrival time of the propagating wavepackets. Using a fiber stretcher in conjunction with a free-space delay line also allows the link to be stabilized at one wavelength while achieving similarly high levels of stabilization for another.

Time multiplexing of the classical and quantum channels adds complexity to the feedback electronics. To prevent the servo from unlocking when the chopper blocks the stabilization light, we use an FPGA-based feedback algorithm \cite{Pomponio2020} where we implement a digital low-pass filter to limit the response bandwidth to less than the chopper speed. This ensures that the servo responds only to the averaged phase error over multiple chopper cycles, rather than to the brief signal dropouts.

\textbf{Optical phase distribution across the network:} For future quantum network experiments, the remaining 96\% of the classical light that is not reflected from the remote end of the fiber for stabilization may be used to distribute a common optical phase across all nodes \cite{Stolk2025,Clivati2022}. Such a phase reference may be used to lock an optical frequency comb~\cite{fortier201920} at each node, used to pump the generation of single-photon states. The comb may also be used to generate a radio frequency signal to clock electronic circuits that is coherent with the distributed optical signal.

%%%%%%%%%%%%%%%%%%%%%%%%%%%%%%%%%%%%%%%%%%%%%%%%%%%%%
\subsection{Quantum channel stability as measured by a classical signal} \label{sec3b}

\begin{figure}
  \centering
  \includegraphics[width=8.5cm]{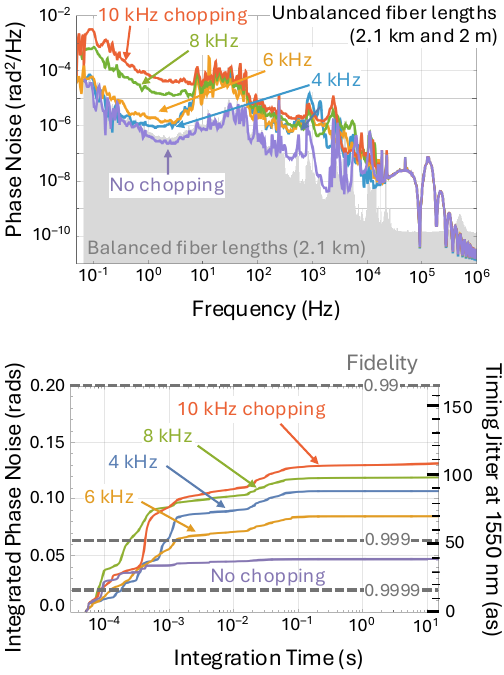}
  \caption{(a) Phase noise power spectral density (PSD) of the 1550 nm optical beatnote between a 2.1 km and 2 m stabilized optical fiber links, showing the frequency-dependent phase fluctuations as a function of chopping rate. The light outputs from the two fibers are heterodyned to enable the combined measurement noise of both links, with the 2 m fiber contributing much less. The gray shaded area shows the noise floor when the two fibers are matched in length. (b) Corresponding integrated root-mean-square (RMS) phase deviation, illustrating the cumulative RMS phase noise as a function of integration time.}
  \label{fig:phase-noise}
\end{figure}

Although the level of frequency stabilization achieved over fiber in atomic clock experiments is impressive, it is not sufficient for transmitting path-entangled states. This is because high-frequency phase fluctuations are averaged away during frequency counting, and do not contribute to the frequency measurement. However, measurements of path entanglement require accurate knowledge of the relative phase between paths and cannot be averaged. As the phase error increases over time due to unobserved path length changes, the fidelity of the entangled state distribution decreases. Consequently, to evaluate how path variations impact quantum state measurement, we must measure the residual optical phase power spectral density, $S_{\phi}(f)$, of the link and integrate this noise from the shortest data acquisition time up to the relevant measurement duration, or integration time, $\tau_{int}$,
\begin{equation}\label{Eq:integrated phase noise}
    \Delta \phi_\mathrm{RMS} = \sqrt{\int_{1/50\mu s}^{1/\tau_{int}} S_{\phi}(f) df}.
\end{equation}

To measure $S_{\phi}(f)$, we independently stabilize two fiber links, one with a length of 2.1 km and the other about 2 m, with the same 1542 nm chopped CW laser. We assume that the longer fiber contributes nearly all of the added phase noise. To classically test the phase noise of the quantum channel, we split and transmit optical pulses derived from a stabilized mode-locked laser through each fiber and interfere the outputs on a high-speed photodetector (1.5 GHz bandwidth) at the fiber outputs. The mode-locked laser is based on a 500 MHz repetition rate Er/Yb:glass oscillator \cite{nardelli2023optical}, whose carrier-envelope offset frequency is stabilized \cite{Telle1999,Reichert1999,Jones2000} and whose spectrum is phase-stabilized to the 1542 nm CW laser. The pulse is filtered by a fiber Bragg grating centered at 1550 nm and with a full-width half-maximum bandwidth of 1 nm.

The goal of the present work is not to improve the performance of earlier fiber link demonstrations~\cite{Ma1994, Williams2008, Clivati2022}, but to provide a stabilized infrastructure with quantum-compatible performance for future path-entanglement and multi-node interference experiments over deployed networks. In contrast to classical time–frequency links that typically aim to maximize fiber length, our approach deliberately limits the link length to achieve a target quantum-state fidelity, as phase noise accumulated across the full Fourier spectrum directly leads to decoherence and reduced interference visibility of path-entangled states. The demonstrated architecture therefore addresses a complementary regime in which high phase stability and indistinguishability are prioritized over absolute distance, enabling scalable quantum networking protocols under realistic coexistence constraints.

In Figure \ref{fig:phase-noise} we perform classical measurements to characterize the quantum channel. The figure shows the measured phase noise power spectral density, and corresponding integrated phase noise for different chopper rates of the 1550 nm quantum channel, while stabilized with the 1542nm classical channel. A slow chopper speed limits the bandwidth of the fiber stretcher feedback and will therefore be inadequate to fully suppress high-bandwidth noise (100 Hz to 2 kHz) that the fiber encodes on the light. This is seen in the 4 kHz phase noise trace (blue) as peaks near 1 kHz, which contribute much of the integrated phase noise. Conversely, a high chopping speed introduces additional noise across all frequencies due to the mechanical timing jitter of the chopping wheel, which increases at higher speeds. This is evident in the red and green phase noise spectra as chopping increases from 4 kHz to 10 kHz, and the effect is absent when the CW laser is not chopped. 

The curves in the top plot of Fig. \ref{fig:phase-noise} are integrated from right to left (short to long timescales) according to Equation \ref{Eq:integrated phase noise} to produce the results in the bottom plot. We set the integration lower bound to 50 $\mu$s (20 kHz), as this is the frequency at which the fiber begins to introduce phase noise. The features above ~20 kHz are delay-line interference fringes in the measured phase-noise spectrum, arising from the effective optical path mismatch between the two arms of the heterodyne measurement, $\tau_{eff}$. This delayed-interference geometry maps the laser’s intrinsic frequency noise into the phase-noise spectrum through the transfer function of a delay-line interferometer, producing periodic spectral replicas at multiples of $1 /\tau_{eff}$\cite{Tsuchida2011}. Because these replicas mask the true high-frequency fiber noise, we do not interpret the spectrum in this region as representative of the link. Instead, our jitter and fidelity calculations rely on the frequency band below the first replica, where the measured phase noise accurately reflects the fiber-induced fluctuations. The gray shaded region in Fig. \ref{fig:phase-noise} demonstrates the lower phase noise (unchopped) when the fiber lengths are matched at 2.1 km. Although this removes the delay-line spectral features, we opted to measure in the unbalanced scenario to isolate the phase noise of a single fiber.

The righthand axis shows the optical timing jitter associated with the integrated phase noise, $\Delta \tau_\mathrm{RMS} = \Delta \phi_\mathrm{RMS}/2\pi f_0$, where $f_0$ is the optical frequency, and theoretical path state fidelity is given in Eq. \ref{Eq:fidelity}.
The fidelity is a measure of the indistinguishability of two states produced when one photon propagates along two paths, according to Eq. \ref{Eq:quantum state}. Temporal relative phase change between two optical paths can degrade the fidelity. 

For fiber length stabilization, in the absence of chopping, we achieve an optical timing jitter well below 50 as, which theoretically maps to a fidelity of $F > 0.999$. Among the different chopper rates, 6 kHz gives the best results with an optical timing jitter of about 70 as, corresponding to $F \approx 0.998$, out to an integration time of at least 20 seconds. Although this is evidence that optical chopping degrades the fidelity, chopping is required for high classical/quantum channel isolation in this demonstration.

The integration time defines how long we can assume a nearly fixed phase relationship between fiber paths, setting the maximum duration over which we can perform projective quantum measurements reliably. To achieve higher state fidelity, measurements can be taken over shorter time windows, with phase recalibration between windows. This approach allows multiple measurements to be cascaded for longer protocols or measurements that require additional averaging.

%%%%%%%%%

\subsubsection{In-loop vs. out-of-loop phase noise analysis}

A careful phase-noise analysis of any stabilized system requires a dedicated measurement channel that is not involved in the feedback loop. In this experiment, each fiber is independently stabilized based on the phase information seen on its \textit{in-loop} photodetector. A separate \textit{out-of-loop} detector, located at the opposite beam-splitter port from the SNSPD, is used to characterize the residual phase noise of the two fiber links together. Importantly, this out-of-loop photodetector is not used in any feedback loop and therefore provides an unbiased measure of the stabilized system performance.

In-loop measurements, such as those obtained from detectors used in the feedback loop, can underestimate the true residual phase noise. This is because feedback suppresses the measured error signal while remaining insensitive to noise sources that are written onto the fiber by the actuator, including technical noise that appears as a valid phase signal, such as detector noise or electrical pickup such as ground loops, as well as phase fluctuations that are not common between the classical stabilization channel and the quantum channel. By contrast, an out-of-loop measurement captures all phase noise that remains on the transmitted optical field, including contributions outside the control bandwidth and noise arising from non-common-mode optical paths.

A particularly stringent out-of-loop test is obtained by independently stabilizing two fiber links and interfering them at a common beam splitter, such that the measured phase noise reflects the combined residual instability of both links. This approach provides a conservative assessment of phase stability and is directly applicable to quantum networking scenarios in which independent stabilized links must maintain mutual phase coherence.

%%%%%%%%%%%%%%%%%%%%%%%%%%%%%%%%%%%%%%%%%%%%%%%%%%%%%
\subsection{Limitations of Phase Stability in Optical Fiber}

Because of the finite speed of light in optical fiber, the achievable phase-noise suppression in an actively stabilized link is fundamentally limited by the fiber’s round-trip propagation time. As the fiber length increases, the feedback bandwidth decreases, since the system must wait for light to travel to the remote end and back before applying a correction. This delay limits both how rapidly phase fluctuations can be suppressed and the overall level of suppression attainable at all Fourier frequencies. Longer fibers also accumulate proportionally more environmental noise, further reducing stabilization performance. To overcome this causal limit, a long link must be segmented into multiple shorter sections, each equipped with its own active stabilization loop.

This limitation is characterized by the following expression \cite{Williams2008}, which yields the theoretical best locked phase noise, $S_\text{locked} (f)$ given the phase noise of the unlocked fiber, $S_{\text{unlocked}} (f)$,
\begin{equation}\label{Eq:theory}
S_\text{locked} (f) \approx \frac{1}{3} \left( 2\pi f \frac{nL}{c} \right)^2 S_\text{unlocked} (f).
\end{equation}
In the expression, $n$ is the fiber refractive index, $L$ is the length of the fiber, $c$ is the speed of light in vacuum and $f$ is the Fourier frequency of the power spectral density. The equation assumes uniformly distributed noise along the fiber, which is a good estimate in many cases.

\begin{figure}
  \centering
  \includegraphics[width=8.5cm]{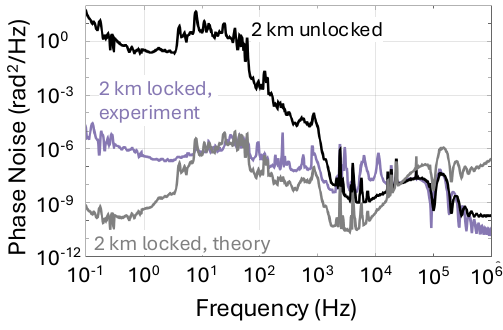}
  \caption{Phase noise power spectral density of the unlocked 2.1 km fiber, $S_\text{unlocked} (f)$ (black), the theoretical best locked fiber $S_\text{locked} (f)$ (gray), and the experimental locked fiber without optical chopping (purple).}
  \label{fig:unlocked-theory}
\end{figure}

In Figure \ref{fig:unlocked-theory} we provides an estimate of the theoretical lower bound on the locked phase noise for our deployed fiber (gray trace) based on the measured unlocked phase noise (black trace) and Equation \ref{Eq:theory}. Note that the assumptions of the equation break down at high frequencies, which manifests as a higher than expected phase noise above 10 kHz in the figure. 

The experimental data for the locked fiber (purple trace) closely matches the theoretical estimate from about 5 Hz to 1 kHz. We also observe that the experimental trace dips below the theoretical trace at certain frequencies, which is possible if the phase noise fluctuations are not perfectly uniform along the length of the fiber. At low frequencies the mismatch is most likely due to differential path length changes between the quantum and classical channels as a result of the difference in wavelengths. At high frequencies (> 1 kHz), the observed servo bump indicates that the loop bandwidth of the fiber stretcher is around 10 kHz, which is not present in the theoretical trace. 

%%%%%%%%%%%%%%%%%%%%%%%%%%%%%%%%%%%%%%%%%%%%%%%%%%%%%
%%%%%%%%%%%%%%%%%%%%%%%%%%%%%%%%%%%%%%%%%%%%%%%%%%%%%
%%%%%%%%%%%%%%%%%%%%%%%%%%%%%%%%%%%%%%%%%%%%%%%%%%%%%
\section{The quantum channel}

To characterize the performance of the quantum channel, we independently stabilize two deployed 2.1 km optical fibers that terminate at the same location (i.e., the same lab). This configuration is used only for phase noise measurement; in a functioning quantum network, the fibers would instead terminate at different locations, linking remote quantum sources. In this measurement, both fibers are part of a 10-fiber bundle, housed in an outdoor-compatible metal jacket, which helps minimize environmentally induced length fluctuations. The bundle is routed through a utility hallway, an environment that introduces more phase noise to the optical signal than a 3.5 km deployed link running across the city of Boulder, CO, making it a good test case for a deployed network. 

The same 500 MHz mode-locked laser used to measure the phase noise of the quantum channel (Section \ref{sec:classical}) is attenuated such that there is an average photon number per pulse of $<$ 1, varied from about 0.001 to 0.9 photons per pulse. This ensures that the interference fringes we measure are due to the quantum coherent state, and that ultimately a single-photon state will be compatible with our stabilization architecture. The relative amplitudes distributed to each fiber are tuned to compensate for different fiber link losses by using a half waveplate and polarizing beam splitter to split the coherent state.

At the other end of the fibers, the quantum and classical channels are split by a 50/50 beam splitter that directs half the amplitude to the final beam splitter where the quantum states from the two network links are combined and interfered. To decrease the channel losses and further isolate the quantum and classical channels, the 50/50 beam splitter can be replaced by a dichroic mirror that splits the 1542 nm and 1550 nm photons. A piezo-driven stage is located in one of the two interferometer arms to slowly change the relative phase of the two quantum channels. 

The photons exiting one of the ports of the final beam splitter are monitored by a superconducting nanowire single-photon detector (SNSPD) with more than 90\% detection efficiency \cite{Reddy2020}. A linear polarizer selects a single polarization to minimize link distinguishability from polarization rotation. A second mechanical chopper is synchronized with the first chopper that chops the CW stabilization laser but with a relative phase of 180 degrees to block the remainder of classical light in the quantum channel. The second chopper has a 25\% duty cycle to increase the robustness of the chopper synchronization at the expense of quantum channel throughput. A 1550 nm narrow bandpass filter serves to further isolate the classical and quantum channels.

\begin{figure}
  \centering
  \includegraphics[width=8.5cm]{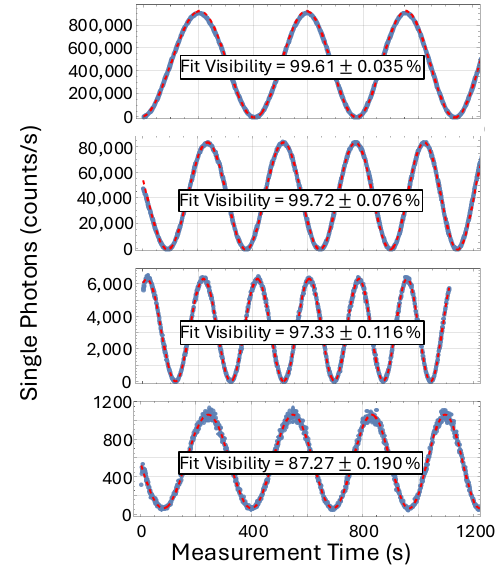}
  \caption{Demonstration of quantum-compatible indistinguishability between two 2.1 km stabilized fibers arranged in a Mach-Zehnder interferometer configuration. A free space optical path length is swept via piezo stage (see Fig. \ref{fig:setup}) while an SNSPD detects single photons out of one of the interferometer ports for four different numbers of average photons. Sinusoidal fits show visibilities of 99.61 $\pm$ 0.035 \% (average of 0.9 photons/pulse), 99.72 $\pm$ 0.076 \% (0.08 photons/pulse), 97.33 $\pm$ 0.116 \% (0.006 photons/pulse) and 87.27 $\pm$ 0.190 \% (0.001 photons/pulse), from top to bottom. The stated uncertainties represent the 95\% confidence intervals. }
  \label{fig:fringes}
\end{figure}

%%%%%%%%%%%%%%%%%%%%%%%%%%%%%%%%%%%%%%%%%%%%%%%%%%%%%
\subsection{Fiber link indistinguishability}

We demonstrate in Fig. \ref{fig:phase-noise} that the network links can support a quantum signal with an integrated phase noise of less than 0.1 rads but it is also necessary that the links be indistinguishable from the point of view of the quantum repeater. This means that the two paths must have the same polarization rotation, attenuation, chromatic dispersion, and the quantum wavepackets must arrive at the same time. 

We stabilize two similar fibers (6 kHz chopping rate) and measure the fiber path indistinguishability by varying the relative phase of the paths with a piezo stage and tracing visibility curves, $V = (\text{max} - \text{min})/(\text{max} + \text{min})$, as seen in Fig. \ref{fig:fringes}. In the figure, we compare the visibilities of four cases with different average numbers of photons per pulse, which we set by attenuating before splitting between the two fiber paths. We fit a sinusoidal signal to each dataset. The highest visibility (more than 99.7\%) is seen in the second case with a peak of about 80,000 counts/second, corresponding to an average of 0.08 photons per pulse. This is limited by the small amount of distinguishability in the two paths. In the third and fourth cases, the average photon number is sufficiently low that the visibility is limited by the SNSPD background counts, which is about 70 counts/s.

The phase delay responsible for sweeping the fringes is driven by a piezo stage, whose displacement has a nonlinear response to voltage. For this is the reason, the fringes in Fig. \ref{fig:fringes} have different temporal periods. To correctly fit a sinusoid, it is necessary to add a chirp parameter to the phase, i.e., $f_{fit}(t) = a \sin(b_3 t^3 + b_2 t^2 + b_1 t + b_0) + c$. However, the high quality of the fit in each case indicates that the two optical paths remain highly indistinguishable throughout the measurement despite the fact that neither polarization nor attenuation is actively controlled.

%%%%%%%%%%%%%%%%%%%%%%%%%%%%%%%%%%%%%%%%%%%%%%%%%%%%%
\subsection{Fiber link stability as measured by a quantum signal}

\begin{figure}
  \centering
  \includegraphics[width=8.5cm]{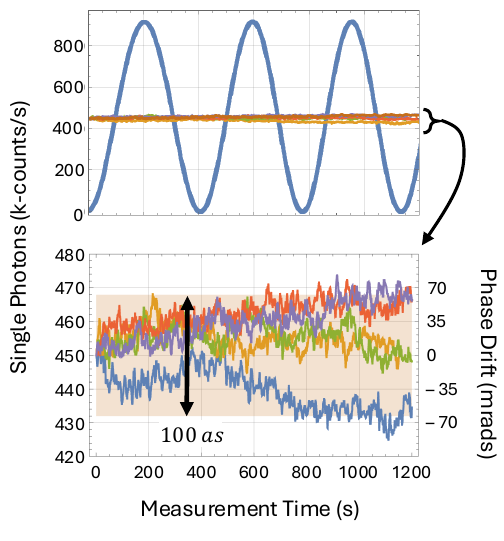}
  \caption{Top: a fringe sweep determines the mid-point where slow phase drift is monitored over five trials. Bottom: the five phase drift trials are shown zoomed in relative to the top plot.}
  \label{fig:phase-wander}
\end{figure}

As we assessed the short-term (< 10 s) phase drift in section \ref{sec:classical}, we assess the long-term phase drift (> 10 s) in this section. Noise in the two regimes is caused by different experimental factors. Fast phase fluctuations (10 Hz to 10 kHz) are primarily caused by the pickup of acoustic and vibrational noise on the fiber link. High-frequency fiber noise is more difficult to suppress and requires a high-bandwidth and a precisely tuned PID feedback algorithm. Slow phase drifts (< 10 Hz) are easy to suppress in feedback, so the presence of this noise in the measurement indicates that the stabilization laser is not completely sampling the quantum channel path. This occurs in part because there are short free-space optical paths where the quantum and classical channels do not co-propagate. Any unsampled path length variations will manifest as phase noise that reduces fidelity. In the fiber itself, different polarizations and different wavelengths experience different phase changes, and so it is possible that our lack of strict polarization control and identical wavelengths for the classical and quantum channels may contribute to the slow phase variation \cite{Clivati2022}.

We measure the slow variation by sweeping the interferometer fringes with the piezo-driven stage and then setting the phase close to the fringe mid-point, which is the steepest and most sensitive to phase variation. As seen in Fig. \ref{fig:phase-wander}, the fringe peak was around 900,000 counts/s so we set the initial phase to about 450,000 counts/s. Over several trials running for 20 minutes each we measure that the slow phase drift consistently stays within the 100 as bounds for about 10 minutes. This does not account for fast phase variation, which is largely averaged away due to the 1 second integration time of the single-photon detector. Additionally, we have not taken into account piezo creep, where the piezo continues to change length for up to hours after being actuated. Therefore, the phase drift presented in this study represents a worse case. 

In the future, this slow phase variation could be further improved by reducing the optical path lengths that are not common to both quantum and classical channels, namely at the central node where the stabilization laser and single photon detector are located. The experiment would also benefit from more passive stabilization, such as employing a custom temperature-stabilized baseplate instead of an optical breadboard.

%%%%%%%%%%%%%%%%%%%%%%%%%%%%%%%%%%%%%%%%%%%%%%%%%%%%%%%%%%%% 
%%%%%%%%%%%%%%%%%%%%%%%%%%%%%%%%%%%%%%%%%%%%%%%%%%%%%%%%%%%% 
%%%%%%%%%%%%%%%%%%%%%%%%%%%%%%%%%%%%%%%%%%%%%%%%%%%%%%%%%%%% 
\section{Isolation of the quantum and classical channels}
\label{sec:isolation}

\begin{figure}
  \centering
  \includegraphics[width=8.5cm]{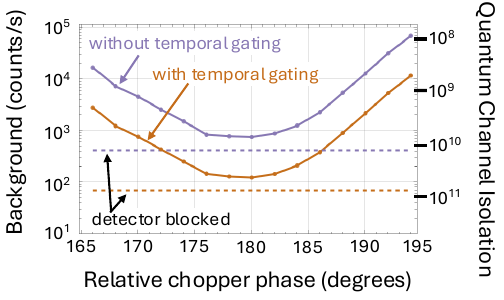}
  \caption{Background counts due to CW stabilization laser and detector counts while blocked as a function of the relative phase between the two choppers. Experiments were done with/without temporal gating in post-processing. The lowest isolation achieved was $(8.7 \pm 1.5) \times 10^{10}$.}
  \label{fig:CW-isolation}
\end{figure}

To fully take advantage of the phase and length stabilization of the classical channel, there must be a very high degree of isolation between the classical channel and the quantum channel. This is because the typical classical channel has many orders-of-magnitude more photons per second than the quantum channel. For example: if the quantum channel contains 1 photon on average for every 100 laser pulses (500 MHz / 100 = 5 million photons/sec), and the classical channel contains 1 mW of power to stabilize the fiber and also the mode-locked laser at the remote node, then an isolation of $10^9$ will ensure that there are equal numbers of photons from the two channels at the single-photon detector. 

Achieving such a high degree of isolation is particularly challenging in the case of optical fiber phase stabilization since the quantum and classical light sources must experience the same path noise. The best stabilization would result from a classical light source that has the same polarization, wavelength and bandwidth as the quantum light source, but quantum/classical isolation requires some distinguishable property. The CW stabilization laser is not actively polarization-controlled after propagation through the deployed fiber. Because the stabilization feedback senses path-length fluctuations independent of polarization, this does not impact phase stabilization. A polarizer is placed only in the quantum detection path to project the single-photon states onto a common polarization mode and remove polarization-induced distinguishability. Distinguishing the two channels based on polarization is difficult because there is strong coupling between orthogonal polarization states in optical fiber and high isolation would require several stages of polarizers.

Isolation based on wavelength \cite{Thomas2024} is more feasible due to high-quality off-the-shelf dichroic mirrors, but this method is fundamentally limited by the nonlinear scattering of classical photons into the quantum band via the Raman effect \cite{Raman1928,Stolen1989}. Even with perfect wavelength discrimination, spontaneous Raman scattering would saturate the quantum channel if distinguished between channels by wavelength only. A 2.1 km silica fiber would scatter 1542 nm light into a 1 nm band around 1550 nm with an efficiency around $10^{-8}$, which would yield an quantum/classical isolation of at most $10^8$ in the experiment \cite{burenkov2023}.

In the present setup, we multiplex the two channels in time with an optical chopper, which, in principle, allows the quantum signal to propagate along a completely dark fiber, eliminating the Raman noise photons, while being periodically stabilized. We demonstrate a classical/quantum channel isolation of $(8.7 \pm 1.5) \times 10^{10}$ by combining the optical chopping approach with an optical filter and a temporal gating technique in post-detection processing. With current channel losses and around 450 $\mu$W in the classical channel, this level of isolation means that 1 out of every 10,000,000 quantum wavepackets will have a CW noise photon.

The optical filter rejects about a factor of 100 of the classical light at 1542 nm, while transmitting the quantum light at 1550 nm. These wavelengths are close enough that they largely experience the same phase fluctuations in fiber, yet far enough that they can be isolated from one another with an optical bandpass filter. Wavelength isolation could be improved by several orders of magnitude with a filter with a steeper band edge.

The temporal gating technique relies on the fact that photons from the quantum channel arrive at precisely defined time intervals because they originate from a pulsed laser, whereas the classical photons are distributed across all times since they originate from a CW laser. The temporal gating technique exploits the fact that photons from the quantum channel arrive at well-defined times determined by the pulsed laser repetition rate, whereas classical photons originating from the CW channel are distributed uniformly in time. Here, “gating” refers to a temporal acceptance window applied to time-tagged detection events. We choose a coincidence window of 250 ps, which equates to a duty cycle of 0.125 over the 2 ns modelocked laser pulse spacing. The window is synchronized to the pulsed source, which reduces both classical leakage counts and detector background counts by approximately an order of magnitude, as shown in Fig. \ref{fig:CW-isolation}. 

\subsection{Chopper duty cycle}
Importantly, the chopper before the SNSPD has a duty cycle of 25\% compared to 50\% for the chopper that chops the CW light. This allows for imperfect chopper synchronization and extra isolation from CW photons that are reflected multiple times through the fiber. However, a lower duty cycle also decreases the quantum channel throughput, reducing the maximum usable rate of the quantum network. One may decrease the duty cycle of the classical channel chopper while increasing the duty cycle of the quantum channel, though this may also decrease the phase stability of the link. This is because a lower duty cycle $D$ reduces the signal-to-noise ratio of the feedback signal and increases detection noise $S_{\phi, det}$ such that $S_{\phi, det} \propto 1 / D$. If we assume a simple phase noise model for the fiber $S_{\phi, fiber} \propto 1/f^2$, then the frequency $f_c$ at which $S_{\phi, det} = S_{\phi, fiber}$ limits the feedback bandwidth. In this case the bandwidth scales with duty cycle as $f_c \propto D^{1/2}$, which yields $\Delta\phi_\mathrm{RMS} \approx \sqrt{S_{\phi, det} \cdot f_c} \propto D^{-1/4}$. If we reduce the CW duty cycle by 10 times from 50\% to 5\%, this corresponds to an increase in integrated phase noise of about 1.8 times. 

Though the back-of-the-envelope calculation indicates quite favorable noise scaling with duty cycle, in practice, reducing the CW duty cycle to 25$\%$ made the phase locked loop unstable and extremely susceptible to unlocking due to small perturbations. Nonetheless, further exploration of the duty cycle and how it relates to loop stability and residual noise would be valuable, though it is beyond the scope of this experiment.

%%%%%%%%%%%%%%%%%%%%%%%%%%%%%%%%%%%%%%%%%%%%%%%%%%%%%%%%%%%% 
%%%%%%%%%%%%%%%%%%%%%%%%%%%%%%%%%%%%%%%%%%%%%%%%%%%%%%%%%%%% 
%%%%%%%%%%%%%%%%%%%%%%%%%%%%%%%%%%%%%%%%%%%%%%%%%%%%%%%%%%%% 
\section{Conclusion and Outlook}

Quantum networks based on path entanglement are promising because they enable high-fidelity quantum state teleportation due to the combination of heralded single photons (for example, from a down-conversion source) with heralded entanglement swapping across remote network nodes. This scheme presents a promising approach for achieving a sufficiently high quantum state fidelity across a lossy network to support protocols with a clear quantum advantage. For the first time, we demonstrate phase-stabilized deployed km-length optical fiber links that are capable of supporting the distribution of path-entangled states with a fidelity greater than 0.99. We have achieved this by using precision frequency control techniques to stabilize optical phase across a fiber link and a novel scheme for time multiplexing classical and quantum signals, demonstrating high channel isolation.

For a future multi-node quantum network, path length phase must be stable not only over a single link but across the entire network of many km-scale links. This will require further suppression of fiber phase noise to guarantee a combined fidelity > 0.99 across all nodes. Passive shielding, including underground fiber conduits, will reduce the high-frequency phase noise that limits the current results. A functional network will also require active stabilization of fiber polarization and loss to ensure link indistinguishability over time. Dispersion will also need to be managed to match the pulse temporal profiles after propagating through network fibers of different lengths. 

Other novel techniques could allow for high quantum/classical isolation without the need to chop, thus reducing the phase noise and channel loss added by the chopper. These include low-loss hollow-core fibers that suffer more than 1000 times less channel crosstalk via Raman scattering \cite{Chen2024}, or multi-core fibers that contain many highly correlated cores in the same cladding, exhibiting a high degree of isolation \cite{hoghooghi2025}. 

\section{Acknowledgment}
The authors acknowledge support from the National Science Foundation QLCI OMA-2016244, National Science Foundation NQVL:QSTD:Pilot 2435378, the National Science Foundation ECCS 2330228, the University of Colorado Quantum Engineering Initiative, and the National Institute of Standards and Technology.

\section{Disclaimer}
Any mention of commercial products within the manuscript is for information only; it does not imply recommendation or endorsement by NIST.

\section{Disclosures}
The authors declare no conflicts of interest.

\section{Data availability}
Data underlying the results presented in this paper are not publicly available at this time but may be obtained from the authors upon reasonable request.

\section*{References}

\bibliography{main.bib}

@PREAMBLE{
 "\providecommand{\noopsort}[1]{}" 
 # "\providecommand{\singleletter}[1]{#1}%" 
}

@ARTICLE{BACON2021,
   author       = "Boulder Atomic Clock Optical Network (BACON) Collaboration",
   title        ="Frequency ratio measurements at 18-digit accuracy using an optical clock network",
   year         = "2021",
   journal      = "Nature",
   volume       = "591",
   pages        = "564–569",
}

@ARTICLE{Ma1994,
   author       = "L.-S. Ma and P. Jungner and J. Ye and J. L. Hall",
   title        ="Delivering the same optical frequency at two places: accurate cancellation of phase noise introduced by an optical fiber or other time-varying path",
   year         = "1994",
   journal      = "Opt. Lett.",
   volume       = "19",
   pages        = "1777-1779",
}

@article{nardelli2023optical,
  title={Optical and microwave metrology at the 10-18 level with an Er/Yb: glass frequency comb},
  author={Nardelli, Nicholas V and Leopardi, Holly and Schibli, Thomas R and Fortier, Tara M},
  journal={Laser \& Photonics Reviews},
  volume={17},
  number={4},
  pages={2200650},
  year={2023},
  publisher={Wiley Online Library}
}

@ARTICLE{Telle1999,
   author       = "H. R. Telle and G. Steinmeyer and A. E. Dunlop and J. Stenger and D. H. Sutter and U. Keller",
   title        = "Carrier-envelope offset phase control: A novel concept for absolute
optical frequency measurement and ultrashort pulse generation",
   year         = "1999",
   journal      = "Appl. Phys. B",
   volume       = "69",
   pages        = " 327–332",
}

@ARTICLE{Reichert1999,
   author       = "J. Reichert and R.Holzwarth amd T. Udem and T. W. Hänsch",
   title        = "Measuring the frequency of light with mode-locked lasers",
   year         = "1999",
   journal      = "Optics Communications",
   volume       = "172",
   pages        = "59-68",
}

@ARTICLE{Pomponio2020,
   author       = "M. Pomponio and A. Hati and C. Nelson",
   title        = "FPGA-based Low-Latency Digital Servo for Optical Physics Experiments",
    journal = "Joint Conference of the IEEE International Frequency Control Symposium and International Symposium on Applications of Ferroelectrics (IFCS-ISAF)",
   pages     = "1-2",
   year      = "2020",
}

@ARTICLE{Matei2017,
   author       = "D. G. Matei and T. Legero and S. Häfner and C. Grebing and R. Weyrich and W. Zhang and L. Sonderhouse and J. M. Robinson and J. Ye and F. Riehle and U. Sterr",
   title        = "1.5 $\mu$m Lasers with Sub-10 mHz Linewidth",
    journal = "Physical Review Letters",
    volume = "118",
   pages     = "263202",
   year      = "2017",
}

@ARTICLE{Jones2000,
   author       = "D.J. Jones and S.A. Diddams and J.K. Ranka and A. Stentz and R.S. Windeler and J.L. Hall and S.T. Cundiff",
   title        = "Carrier-envelope phase control of femtosecond mode-locked lasers and direct optical frequency synthesis",
   year         = "2000",
   journal      = "Science",
   volume       = "288",
   pages        = "635-639",
}

@article{fortier201920,
  title={20 years of developments in optical frequency comb technology and applications},
  author={Fortier, Tara and Baumann, Esther},
  journal={Communications Physics},
  volume={2},
  number={1},
  pages={1--16},
  year={2019},
  publisher={Nature Publishing Group}
}

@article{burenkov2023,
  title={Synchronization and coexistence in quantum networks},
  author={Burenkov, Ivan A and Semionov, Alexandra and Hala and Gerrits, Thomas and Rahmouni, Anouar and Anand, DJ and Li-Baboud, Ya-Shian and Slattery, Oliver and Battou, Abdella and Polyakov, Sergey V},
  journal={Optics Express},
  volume={31},
  number={7},
  pages={11431--11446},
  year={2023},
  publisher={Optica Publishing Group}
}

@article{hoghooghi2025,
  title={Ultrastable optical frequency transfer and attosecond timing in deployed multicore fiber},
  author={Hoghooghi, Nazanin and Mazur, Mikael and Fontaine, Nicolas and Liu, Yifan and Lee, Dahyeon and McLemore, Charles and Nakamura, Takuma and Hayashi, Tetsuya and Di Sciullo, Giammarco and Shaji, Divya and others},
  journal={Optica},
  volume={12},
  number={6},
  pages={894--899},
  year={2025},
  publisher={Optica Publishing Group}
}

@article{Reddy2020,
	title = {Superconducting nanowire single-photon detectors with 98\% system detection efficiency at 1550 nm},
	volume = {7},
	issn = {2334-2536},
	url = {https://opg.optica.org/optica/abstract.cfm?uri=optica-7-12-1649},
	doi = {10.1364/OPTICA.400751},
	number = {12},
	urldate = {2025-02-25},
	journal = {Optica},
	author = {Reddy, Dileep V. and Nerem, Robert R. and Nam, Sae Woo and Mirin, Richard P. and Verma, Varun B.},
	month = dec,
	year = {2020},
	note = {Publisher: Optica Publishing Group},
	pages = {1649--1653},
}

@article{DLCZ2001,
  title={Long-distance quantum communication with atomic ensembles and linear optics},
  author={Duan, L-M and Lukin, Mikhail D and Cirac, J Ignacio and Zoller, Peter},
  journal={Nature},
  volume={414},
  number={6862},
  pages={413--418},
  year={2001},
  publisher={Nature Publishing Group UK London}
}

@article{Gottesman2012,
  title={Longer-baseline telescopes using quantum repeaters},
  author={Gottesman, Daniel and Jennewein, Thomas and Croke, Sarah},
  journal={Physical review letters},
  volume={109},
  number={7},
  pages={070503},
  year={2012},
  publisher={APS}
}

@article{Czupryniak2023,
  title={Optimal qubit circuits for quantum-enhanced telescopes},
  author={Czupryniak, Robert and Steinmetz, John and Kwiat, Paul G and Jordan, Andrew N},
  journal={Physical Review A},
  volume={108},
  number={5},
  pages={052408},
  year={2023},
  publisher={APS}
}

@article{Zhou2023,
  title={Twin-field quantum key distribution without optical frequency dissemination},
  author={Zhou, Lai and Lin, Jinping and Jing, Yumang and Yuan, Zhiliang},
  journal={nature communications},
  volume={14},
  number={1},
  pages={928},
  year={2023},
  publisher={Nature Publishing Group UK London}
}

@article{Lucamarini2018,
  title={Overcoming the rate--distance limit of quantum key distribution without quantum repeaters},
  author={Lucamarini, Marco and Yuan, Zhiliang L and Dynes, James F and Shields, Andrew J},
  journal={Nature},
  volume={557},
  number={7705},
  pages={400--403},
  year={2018},
  publisher={Nature Publishing Group UK London}
}

@article{Williams2008,
  title={High-stability transfer of an optical frequency over long fiber-optic links},
  author={Williams, Paul A and Swann, William C and Newbury, Nathan R},
  journal={Journal of the Optical Society of America B},
  volume={25},
  number={8},
  pages={1284--1293},
  year={2008},
  publisher={Optical Society of America}
}

@article{Raman1928,
  title={The negative absorption of radiation},
  author={Raman, CV and Krishnan, KS},
  journal={Nature},
  volume={122},
  number={3062},
  pages={12--13},
  year={1928},
  publisher={Nature Publishing Group UK London}
}

@article{Stolen1989,
  title={Raman response function of silica-core fibers},
  author={Stolen, Roger H and Gordon, James P and Tomlinson, WJ and Haus, Hermann A},
  journal={Journal of the Optical Society of America B},
  volume={6},
  number={6},
  pages={1159--1166},
  year={1989},
  publisher={Optical Society of America}
}

@inproceedings{Chen2024,
  title={Hollow core DNANF optical fiber with< 0.11 dB/km loss},
  author={Chen, Y and Petrovich, MN and Fokoua, E Numkam and Adamu, AI and Hassan, MRA and Sakr, H and Slav{\'\i}k, R and Gorajoobi, S Bakhtiari and Alonso, M and Ando, R Fatobene and others},
  booktitle={Optical Fiber Communication Conference},
  pages={Th4A--8},
  year={2024},
  organization={Optica Publishing Group}
}

@article{Thomas2024,
  title={Quantum teleportation coexisting with classical communications in optical fiber},
  author={Thomas, Jordan M and Yeh, Fei I and Chen, Jim Hao and Mambretti, Joe J and Kohlert, Scott J and Kanter, Gregory S and Kumar, Prem},
  journal={Optica},
  volume={11},
  number={12},
  pages={1700--1707},
  year={2024},
  publisher={Optica Publishing Group}
}

@article{Clivati2022,
  title={Coherent phase transfer for real-world twin-field quantum key distribution},
  author={Clivati, Cecilia and Meda, Alice and Donadello, Simone and Virz{\`\i}, Salvatore and Genovese, Marco and Levi, Filippo and Mura, Alberto and Pittaluga, Mirko and Yuan, Zhiliang and Shields, Andrew J and others},
  journal={Nature communications},
  volume={13},
  number={1},
  pages={157},
  year={2022},
  publisher={Nature Publishing Group UK London}
}

@article{Tsuchida2011,
  title={Laser frequency modulation noise measurement by recirculating delayed self-heterodyne method},
  author={Tsuchida, Hidemi},
  journal={Optics letters},
  volume={36},
  number={5},
  pages={681--683},
  year={2011},
  publisher={Optical Society of America}
}

@article{Stolk2025,
  title={Extendable optical phase synchronization of remote and independent quantum network nodes over deployed fibers},
  author={Stolk, AJ and Biemond, JJB and van der Enden, KL and van Dooren, L and van Zwet, EJ and Hanson, R},
  journal={Physical Review Applied},
  volume={23},
  number={1},
  pages={014077},
  year={2025},
  publisher={APS}
}

@article{Liu2024,
  title={Creation of memory--memory entanglement in a metropolitan quantum network},
  author={Liu, Jian-Long and Luo, Xi-Yu and Yu, Yong and Wang, Chao-Yang and Wang, Bin and Hu, Yi and Li, Jun and Zheng, Ming-Yang and Yao, Bo and Yan, Zi and others},
  journal={Nature},
  volume={629},
  number={8012},
  pages={579--585},
  year={2024},
  publisher={Nature Publishing Group UK London}
}

@article{PhysRevA.76.062323,
  title = {Distributed quantum computation based on small quantum registers},
  author = {Jiang, Liang and Taylor, Jacob M. and S\o{}rensen, Anders S. and Lukin, Mikhail D.},
  journal = {Phys. Rev. A},
  volume = {76},
  issue = {6},
  pages = {062323},
  numpages = {22},
  year = {2007},
  month = {Dec},
  publisher = {American Physical Society},
  doi = {10.1103/PhysRevA.76.062323},
  url = {https://link.aps.org/doi/10.1103/PhysRevA.76.062323}
}

@article{RevModPhys.74.145,
  title = {Quantum cryptography},
  author = {Gisin, Nicolas and Ribordy, Gr\'egoire and Tittel, Wolfgang and Zbinden, Hugo},
  journal = {Rev. Mod. Phys.},
  volume = {74},
  issue = {1},
  pages = {145--195},
  numpages = {0},
  year = {2002},
  month = {Mar},
  publisher = {American Physical Society},
  doi = {10.1103/RevModPhys.74.145},
  url = {https://link.aps.org/doi/10.1103/RevModPhys.74.145}
}

@article{PhysRevA.59.169,
  title = {Quantum repeaters based on entanglement purification},
  author = {D\"ur, W. and Briegel, H.-J. and Cirac, J. I. and Zoller, P.},
  journal = {Phys. Rev. A},
  volume = {59},
  issue = {1},
  pages = {169--181},
  numpages = {0},
  year = {1999},
  month = {Jan},
  publisher = {American Physical Society},
  doi = {10.1103/PhysRevA.59.169},
  url = {https://link.aps.org/doi/10.1103/PhysRevA.59.169}
}

@article{PhysRevLett.81.5932,
  title = {Quantum Repeaters: The Role of Imperfect Local Operations in Quantum Communication},
  author = {Briegel, H.-J. and D\"ur, W. and Cirac, J. I. and Zoller, P.},
  journal = {Phys. Rev. Lett.},
  volume = {81},
  issue = {26},
  pages = {5932--5935},
  numpages = {0},
  year = {1998},
  month = {Dec},
  publisher = {American Physical Society},
  doi = {10.1103/PhysRevLett.81.5932},
  url = {https://link.aps.org/doi/10.1103/PhysRevLett.81.5932}}

@article{RevModPhys.83.33,
  title = {Quantum repeaters based on atomic ensembles and linear optics},
  author = {Sangouard, Nicolas and Simon, Christoph and de Riedmatten, Hugues and Gisin, Nicolas},
  journal = {Rev. Mod. Phys.},
  volume = {83},
  issue = {1},
  pages = {33--80},
  numpages = {0},
  year = {2011},
  month = {Mar},
  publisher = {American Physical Society},
  doi = {10.1103/RevModPhys.83.33},
  url = {https://link.aps.org/doi/10.1103/RevModPhys.83.33}}

@article{PARIS199678,
	author = {Matteo G.A. Paris},
	doi = {https://doi.org/10.1016/0375-9601(96)00339-8},
	issn = {0375-9601},
	journal = {Physics Letters A},
	number = {2},
	pages = {78-80},
	title = {Displacement operator by beam splitter},
	url = {https://www.sciencedirect.com/science/article/pii/0375960196003398},
	volume = {217},
	year = {1996}}

@article{PhysRevLett.82.2009,
  title = {Testing Quantum Nonlocality in Phase Space},
  author = {Banaszek, Konrad and W\'odkiewicz, Krzysztof},
  journal = {Phys. Rev. Lett.},
  volume = {82},
  issue = {10},
  pages = {2009--2013},
  numpages = {0},
  year = {1999},
  month = {Mar},
  publisher = {American Physical Society},
  doi = {10.1103/PhysRevLett.82.2009},
  url = {https://link.aps.org/doi/10.1103/PhysRevLett.82.2009}
}

@article{PhysRevA.88.012111,
  title = {Testing nonlocality of a single photon without a shared reference frame},
  author = {Brask, Jonatan Bohr and Chaves, Rafael and Brunner, Nicolas},
  journal = {Phys. Rev. A},
  volume = {88},
  issue = {1},
  pages = {012111},
  numpages = {5},
  year = {2013},
  month = {Jul},
  publisher = {American Physical Society},
  doi = {10.1103/PhysRevA.88.012111},
  url = {https://link.aps.org/doi/10.1103/PhysRevA.88.012111}
}

@article{PhysRevA.90.033836,
  title = {Linear optics schemes for entanglement distribution with realistic single-photon sources},
  author = {Lasota, Miko\l{}aj and Radzewicz, Czes\l{}aw and Banaszek, Konrad and Thew, Rob},
  journal = {Phys. Rev. A},
  volume = {90},
  issue = {3},
  pages = {033836},
  numpages = {11},
  year = {2014},
  month = {Sep},
  publisher = {American Physical Society},
  doi = {10.1103/PhysRevA.90.033836},
  url = {https://link.aps.org/doi/10.1103/PhysRevA.90.033836}
}

@article{PhysRevA.76.050301,
  title = {Long-distance entanglement distribution with single-photon sources},
  author = {Sangouard, Nicolas and Simon, Christoph and Min\'a\ifmmode \check{r}\else \v{r}\fi{}, Ji\ifmmode \check{r}\else \v{r}\fi{}\'{\i} and Zbinden, Hugo and de Riedmatten, Hugues and Gisin, Nicolas},
  journal = {Phys. Rev. A},
  volume = {76},
  issue = {5},
  pages = {050301},
  numpages = {4},
  year = {2007},
  month = {Nov},
  publisher = {American Physical Society},
  doi = {10.1103/PhysRevA.76.050301},
  url = {https://link.aps.org/doi/10.1103/PhysRevA.76.050301}
}

@article{PhysRevA.76.052101,
  title = {Strong violations of Bell-type inequalities for path-entangled number states},
  author = {Wildfeuer, Christoph F. and Lund, Austin P. and Dowling, Jonathan P.},
  journal = {Phys. Rev. A},
  volume = {76},
  issue = {5},
  pages = {052101},
  numpages = {6},
  year = {2007},
  month = {Nov},
  publisher = {American Physical Society},
  doi = {10.1103/PhysRevA.76.052101},
  url = {https://link.aps.org/doi/10.1103/PhysRevA.76.052101}
}

@article{riedmantten2021,
	author = {Lago-Rivera, Dario and Grandi, Samuele and Rakonjac, Jelena V. and Seri, Alessandro and de Riedmatten, Hugues},
	date = {2021/06/01},
	doi = {10.1038/s41586-021-03481-8},
	id = {Lago-Rivera2021},
	isbn = {1476-4687},
	journal = {Nature},
	number = {7861},
	pages = {37--40},
	title = {Telecom-heralded entanglement between multimode solid-state quantum memories},
	url = {https://doi.org/10.1038/s41586-021-03481-8},
	volume = {594},
	year = {2021}}

@article{Kaneda:17,
	author = {Fumihiro Kaneda and Feihu Xu and Joseph Chapman and Paul G. Kwiat},
	doi = {10.1364/OPTICA.4.001034},
	journal = {Optica},
	month = {Sep},
	number = {9},
	pages = {1034--1037},
	publisher = {Optica Publishing Group},
	title = {Quantum-memory-assisted multi-photon generation for efficient quantum information processing},
	url = {https://opg.optica.org/optica/abstract.cfm?URI=optica-4-9-1034},
	volume = {4},
	year = {2017}}

@article{Stolk2024,
  title={Metropolitan-scale heralded entanglement of solid-state qubits},
  author={Stolk, Arian J and van der Enden, Kian L and Slater, Marie-Christine and te Raa-Derckx, Ingmar and Botma, Pieter and Van Rantwijk, Joris and Biemond, JJ Benjamin and Hagen, Ronald AJ and Herfst, Rodolf W and Koek, Wouter D and others},
  journal={Science advances},
  volume={10},
  number={44},
  pages={eadp6442},
  year={2024},
  publisher={American Association for the Advancement of Science}
}

\end{document}